\documentclass[aps,pre,print,groupedaddress]{revtex4-1}
\usepackage{graphicx}  
\usepackage{dcolumn}  
\usepackage{bm}           
\usepackage{amsmath}
\usepackage{amsfonts}
\usepackage{amssymb}
\usepackage{color}
\usepackage[normalem]{ulem}
\usepackage{ wasysym }
\usepackage{dsfont}
\usepackage{mathtools}
\usepackage{amsthm}

\usepackage{array}
\newcolumntype{L}[1]{>{\raggedright\let\newline\\\arraybackslash\hspace{0pt}}m{#1}}
\newcolumntype{C}[1]{>{\centering\let\newline\\\arraybackslash\hspace{0pt}}m{#1}}
\newcolumntype{R}[1]{>{\raggedleft\let\newline\\\arraybackslash\hspace{0pt}}m{#1}}

\newtagform{AArabic}[]()

\usepackage{todonotes}

\theoremstyle{definition}

\usepackage{mdframed}
\newmdtheoremenv[style=MyFrame]{teorema}{Result}
\newmdtheoremenv[style=MyFrame]{estrategia}{Strategy}
\newmdtheoremenv[style=MyFrame]{nota}{Remark}
\newmdtheoremenv[style=MyFrame]{consecuencia}{Consequence}
\newmdtheoremenv[style=MyFrame]{definicion}{Definition}

\mdfdefinestyle{MyFrame}{%
    linewidth=0.4mm,
    linecolor=black,
    outerlinewidth=2pt,
    roundcorner=20pt,
    innertopmargin=4pt,
    innerbottommargin=4pt,
    innerrightmargin=4pt,
    innerleftmargin=4pt,
        leftmargin = 4pt,
        rightmargin = 4pt,
        backgroundcolor=gray!50!white
        }

\begin{document}

\title{Turing patterns on a two-component isotropic growing system.\\ Part 3: Time dependent conditions}

\author{Aldo Ledesma-Dur\'an}
\email{aldo\_ledesma@xanum.uam.mx}
\affiliation{UAM-Iztapalapa}

\date{30.~September 2023}

\begin{abstract}
We propose general conditions for the emergence of Turing patterns in a domain that changes size through homogeneous growth/shrinkage based on the qualitative changes of a potential function. For this part of the work, we consider the most general case where the homogeneous state of the system depends on time. Our hypotheses for the Turing conditions are corroborated with numerical simulations of increasing/decreasing domains of the Brusselator system for the linear growth/shrinking case. The simulations allow us to understand the characteristics of the pattern, its amplitude, and wave number, in addition to allowing us to glimpse the role of time as a bifurcation parameter.
\end{abstract}


\maketitle


\section{Presentation}\label{sec:presentation}

The Turing bifurcation in reaction-diffusion systems where the domain changes size is an essential model for understanding patterns in biological systems where, in most cases, the system changes size due to development. We know that the shape of the pattern at a specific time in a growing domain crucially depends on its past history \cite{krause2019influence,klika2017history}. But this is not an exclusive property of growing domains, but the same dependence on past history also occurs in fixed domains. This phenomenon is probably related to persistence, \emph{i.e.} the ability of a dissipative structure to maintain its current wavenumber \cite{ledesma2020eckhaus}. This type of stability in a fixed-size diffusion reaction is known as Eckhaus stability, and its proof requires a nonlinear approximation to the solution of the system near the Turing bifurcation. However, in the case of a domain that changes over time, this analysis is not yet practicable since it has not been conclusively resolved, even from the linear approach, how to find the Turing bifurcation. One of the main problems in this direction is the temporal dependence of homogeneous states.

For this part of the work, we find the Turing bifurcation by considering the changes in the structure of a potential function for perturbations of the Fourier modes. From this potential function, we expect that all trajectories will decay to a stable point in the absence of diffusion, and that some will become (unstable) saddles  for some wavenumber when diffusion is turned on, as we have done in the last part of the work, but now for the more general case where the homogeneous state depends on time. This will establish hypotheses of Turing pattern formation that will be tested against specific numerical simulations of the Brusselator RDD system using the finite difference method in a one-dimensional reaction diffusion system with homogeneous linear growth/shrinkage.

\subsection{Summary of Parts 1 and 2: Time dependent homogeneous state and potential function}\label{sec:antecedents}

In an isotropically growing system where a reaction-diffusion process occurs, the equations describing the dynamic is 
\begin{equation}\label{eq:system}
 \frac{\partial  \mathbf{c}}{ \partial t}  +   \frac{\dot{l}(t)}{l(t)} \mathbf{c}(\xi,t) = \frac{1}{l^2(t)}   \mathds{D}  \frac{ \partial^2  \mathbf{c}}{ \partial \xi^2}  +\mathbf{f}( \mathbf{c}).
 \end{equation}
Here $l(t)$ is the function measuring the expansion/shrinking of the domain, and the relationship between the real and computational domain is $ x =x_0 +l(t) \xi$ with $\xi \in [0, 1]$, where $\xi$ the fixed coordinate and $x$ the actual coordinate. Besides $\mathbf{c}$ represents the concentrations, $\mathds{D} $ the square diffusion matrix, and $\mathbf{f}( \mathbf{c})$ the vector of chemical reactions.

Eq. \eqref{eq:system} can be separated into that for the homogeneous state and for the perturbations, where the latter are assumed to depend on the spatial coordinate (unlike the former) and are comparatively small. The homogeneous state $\mathbf{c}_s(t)$ obeys

\begin{equation}\label{eq:homogeneous}
 \frac{\partial   \mathbf{c}_s}{ \partial t} +\frac{l'(t)}{l(t)} \mathbf{c}_s=\mathbf{f}(\mathbf{c}_s),
\end{equation} 
and in general it depends on the time. In Part 1 \cite{ledesma2023turingA}, we show that under appropriate approximations related to 1) the slow variation of the domain, 2) a sufficient distance of the bifurcations and 3) smallness of nonlinear terms, we show that a good approximation for the homogeneous state satisfying $\mathbf {c} _s(0)=\mathbf{c}_0$ in \eqref{eq:homogeneous} is given by

\begin{equation}\label{eq:homoanalitic}
 \mathbf{c}_s(t)= \mathbf{c}_0-
 \mathds{P} \frac{e^{\boldsymbol{\Lambda} t}}{l(t)} \left(\int \limits_0^t  l'(t') e^{-\boldsymbol{\Lambda} t'}dt' \right)\mathds{P}^{-1}  \mathbf{c}_0.
\end{equation}
Here, $\boldsymbol{\Lambda}$ is the diagonal matrix of eigenvalues of the Jacobian $\mathds{J}\equiv \frac{\partial f}{\partial \mathbf{c}}(\mathbf{c}_0)$, $\mathds{P}$ its modal matrix and $\mathbf{c}_0$ the constant fixed point of the reaction where $\mathbf{f}(c_0)=\mathbf{0}$. 

In constrast,  the perturbations $\boldsymbol{\zeta}$ of the system \eqref{eq:system}, to first order obey 
\begin{equation}\label{eq:perturbations}
 \frac{\partial \boldsymbol{\zeta}}{ \partial t}   +\frac{\dot{l}(t)}{l(t)}\boldsymbol{\zeta} = \frac{1}{l^2(t)}   \mathds{D}  \frac{ \partial^2  \boldsymbol{\zeta}}{ \partial \xi^2}  + \frac{\partial \mathbf{f}}{\partial \mathbf{c}}(\mathbf{c}_s) \boldsymbol{\zeta}.
\end{equation} 
For a two component system, the involved matrices are 

\begin{equation}\label{eq:matrices}
\hat{\mathds{J}}\equiv  \frac{\partial \mathbf{f}}{\partial \mathbf{c}}(\mathbf{c}_s) =\left(\begin{array}{cc}
\hat{j}_{11} & \hat{j}_{12} \\ 
\hat{j}_{21} & \hat{j}_{22}
\end{array}\right)   \mbox{ and }  
\mathds{D}=\left(\begin{array}{cc}
d_{u} & 0\\ 
0 & d_v
\end{array}\right).
\end{equation}
Therefore, the evaluation of the last term in \eqref{eq:perturbations} generally depends explicitly on the time-dependent homogeneous state through the factor $\hat{\mathds{J}}(t)$. 

In Part 2 of this series, we study the Turing conditions under the constant $\mathbf{c}_s\approx $ approximation, which is appropriate, for example, for exponential growth/shrinkage \cite{ledesma2023turingB}. In this case, after taking the Fourier series in the computational domain, for each wavenumber $\kappa$, the Fourier modes obey

\begin{equation} \label{eq:modefourier}
 \frac{\partial \boldsymbol{\zeta}_\kappa}{ \partial t}  = \left[ \hat{ \mathds{J}}-\left(\frac{\kappa}{l(t)}\right)^2\mathds{D} -\frac{\dot{l}(t)}{l(t)} \mathds{I} \right] \boldsymbol{\zeta}_\kappa,
\end{equation}
Here $k(t) \equiv \kappa/l(t)$ is the wave number in the actual domain, and $g(t)\equiv \dot{l}(t)/l(t)$ represents the percentage of size increased/decreased per unit of time. We will call the matrix in parentheses $A(\kappa,t)$.

By first rewriting these equations in \eqref{eq:modefourier} as a pair of second-order equations, we can also write them in potential function form such that $d V_\kappa/dt=F(u_\kappa,u'_\kappa ) $. The qualitative changes of this function allow us to establish the stability properties of the trajectories of each Fourier mode. This allows us to show that the system in the absence of diffusion is stable if

\begin{equation}\label{eq:constant1}
 \Delta_A (0,t)+g'(t) \geq 0 \mbox{ , } \tau_A(0,t) \leq 0 \mbox{ and }  \frac{d}{dt} \left[ \Delta_A(0,t) +g'(t) \right]  \leq 0.
\end{equation}
These conditions guarantee that $V_0$ (the potential function associated with the mode $\kappa=0$) is an elliptical paraboloid centered at the origin, toward which all trajectories are directed. Instability with diffusion requires that for some wavenumber $\kappa_m$, $V_{m} \geq 0$ and that $V_m$ be a saddle, which requires

\begin{equation}\label{eq:constant2}
 \Delta_{A}(\kappa_m,t) +2 d_u k_m(t) k_m'(t)  +g'(t)    <0. 
\end{equation}
The same condition applies also by also using $d_v$ instead of $d_u$. Here and from now on, $\tau$ and $\Delta$ refer to the trace and determination of the matrix in the subscript.

In terms of the original matrices of the RDD system, these conditions are sumarized in the second column of Table \eqref{tab:conditions}. Here $d_i$ refers interchangeably to either $d_u$ or $d_v$. These conditions were applied for example for exponential growth, $l(t)=l(0)e^{rt}$ and their predictions were corroborated against numerical simulations of the Brusselator giving excellent results in the prediction of the Turing space and the number of wave when $|r|<0.1$, and a good prediction of Turing space asymmetries with respect to the Turing and Hopf bifurcations for growth/shrinkage processes \cite{ledesma2023turingB}.

 \section{Turing conditions for growing domain with time dependent homogeneous state}\label{sec:timedependent}

We now consider the more general case where $\mathbf{c}_s$ depends on time and therefore also $ \frac{\partial \mathbf{f}(\mathbf{c}_s)}{\partial \mathbf{c }} = \hat{\mathds{J}}(t)$. After taking the Fourier transform in the computational domain, for each wavenumber $\kappa$, eq. \eqref{eq:system} becomes

\begin{equation}
 \frac{\partial \boldsymbol{\zeta}_\kappa}{ \partial t}  = \left[ \hat{ \mathds{J}}(t)-\left(\frac{\kappa}{l(t)}\right)^2\mathds{D} -\frac{\dot{l}(t)}{l(t)} \mathds{I} \right] \boldsymbol{\zeta}_\kappa,
\end{equation}
and the matrix in parenthesis is now
\begin{equation}\label{eq:caligraphcA}
A(\kappa,t)= \hat{ \mathds{J}}(t)-k^2(t) \mathds{D} -g(t) \mathds{I}. 
\end{equation}

 If we define $\boldsymbol{\zeta}_\kappa=(u_\kappa,v_\kappa)$, the system in component form is 

\begin{align*}
u'_\kappa(t)&=-\left(\frac{\kappa}{l(t)}\right)^2 d_1 u_\kappa + \hat{j}_{11}(t) u + \hat{j}_{12}(t) v_\kappa-g(t) u_k,\\
v'_\kappa(t)&=-\left(\frac{\kappa}{l(t)}\right)^2  d_2 v_\kappa + \hat{j}_{21}(t) u_\kappa + \hat{j}_{22}(t) v_\kappa-g(t) v_k.\\
\end{align*}
The second order equation for the first component  is 

\begin{equation}\label{eq:uequation}
u''_\kappa(t)-\left[\tau_A+ \frac{d}{dt} \log[\hat{j}_{12}] \right] u'_\kappa+ \left[ \Delta_A  +2 d_u k(t) k'(t) +g'(t)-\hat{j}'_{11}+A_{11} \frac{d}{dt} \log[\hat{j}_{12}]  \right] u_\kappa=0,
\end{equation}
and one similar for $v_k$ by changing $d_u \to d_v$, $\hat{j}_{12} \to \hat{j}_{21}$, $\hat{j}'_{11} \to \hat{j}'_{22}$ and $A_{11} \to A_{22}$. Multiplying by $u'_{\kappa}$ and rearranging, we have 

\begin{equation}
\frac{dV_\kappa}{dt} =\left[\tau_A(\kappa)+ \frac{d}{dt} \log[\hat{j}_{12}]  \right]   u_\kappa'^2+ \frac{u^2_\kappa}{2} \frac{d}{dt} \left[  \Delta_A(\kappa)+2 d_u k(t) k'(t) +g'(t) -\hat{j}'_{11}+A_{11}(\kappa) \frac{d}{dt} \log[\hat{j}_{12}]   \right] .
\end{equation}
where the potential function is 

\begin{equation}
V_{\kappa}=\frac{u_\kappa'^2}{2} + \left[ \Delta_A(\kappa) +2 d_u k(t) k'(t)  + g'(t) -\hat{j}'_{11}+A_{11}(\kappa) \frac{d}{dt} \log[\hat{j}_{12}] \right] \frac{u_\kappa^2}{2}.
\end{equation}

The stability conditions in the absence of diffusion ($\kappa=0$) require that $V_0\geq 0$ and $\dot{V}_0 \leq 0$. This implies that

\begin{eqnarray}\label{eq:constant1}
\Delta_A(0) + g'(t) -\hat{j}'_{11}-A_{11}(0) \frac{d}{dt} \log[\hat{j}_{12}] && \geq 0, \\
\tau_A(0)+ \frac{d}{dt} \log[\hat{j}_{12}] && \leq 0, \\
\frac{d}{dt} \left[ \Delta_A(0)  +g'(t) -\hat{j}'_{11}-A_{11}(0) \frac{d}{dt} \log[\hat{j}_{12}]  \right] &&  \leq 0.
\end{eqnarray}
These conditions guarantee that $V_0$ is an elliptical paraboloid centered at the origin toward which all trajectories are directed.

Instability with diffusion requires that for some $\kappa_m$, $V_{m} \geq 0$ and that $V_m$ be a saddle. Therefore

\begin{equation}\label{eq:constant2}
 \Delta_A(\kappa_m) +2 d_u k_m(t) k_m'(t)  + g'(t) -\hat{j}_{11}-A_{11}(\kappa_m) \frac{d}{dt} \log[\hat{j}_{12}] <0
\end{equation}
The condition for $V_m$ to be a potential is that all trajectories descend, $\dot{V}_{m} \leq 0$.

Following the same procedure as in Part 2 of this series, the conditions for the formation of Turing patterns derived in this section in terms of the original matrices are given in Table \ref{tab:conditions}. In the left column we summarize the conditions for a system with a constant homogeneous state \cite{ledesma2023turingB}, and in the right column we add the correction due to the change in time of the homogeneous state found in this work. As in Part 2, we have added the labels $S$, $I$, and $D$, which denote stability, instability, and domain conditions, respectively.

\begin{table*}[h!]
\begin{tabular}{|C{1cm}|C{7cm}|C{7cm}|C{1cm}|}
\hline 
\# & Constant HS  & Correction for Time Dependent HS & -- \\ 
\hline 
\hline
S1) & $\Delta_\mathds{\hat{J}} -\tau_{\mathds{\hat{J}}} g(t)+ g^2(t)+g'(t)  $  & $+\frac{(\hat{j}_{11}(t)-g(t)) \hat{j}_{12}'(t)}{\hat{j}_{12}(t)}-\hat{j}_{11}'(t) $& $>0$\\ 
\hline 
S2)  &  $\tau_\mathds{\hat{J}}-2g(t)$  &$+\frac{\hat{j}_{12}'(t)}{\hat{j}_{12}(t)}$ & $<0$\\ 
\hline 
D3) &  $g'(t) [2g(t)-\tau_\mathds{\hat{J}}]+g''(t) $ & $+\Delta'_{\hat{\mathds{J}}}-\tau'_{\hat{\mathds{J}}}g(t)-\hat{j}_{11}''(t)-\frac{\partial }{\partial t}\left[\frac{(g(t)-\hat{j}_{11}(t)) \hat{j}_{12}'(t)}{\hat{j}_{12}(t)}\right]$ &$\leq 0$ \\ 
\hline
\hline 
I4) & $(2 d_i-\tau_{\mathds{D}}) g(t) + \sigma_{\mathds{D\hat{J}}}$ & + $d_u \frac{d}{dt} \log[\hat{j}_{12}]$ & $>0$ \\ 
\hline 
I5)  & $\sigma_{\mathds{D\hat{J}}}^2- 4 \Delta_{\mathds{D}}\Delta_\mathds{\hat{J}} +2g(t)[ (2 d_j-\tau_{D})\sigma_{\mathds{D\hat{J}}} +2  \tau_{\mathds{\hat{J}}}\Delta_\mathds{D}]$
& $+\frac{\hat{j}_{12}'(t) (2 (d_u \tau_{\mathds{D}} g(t)+d_u \text{$\sigma $0}-2 \Delta_{\mathds{D}} \hat{j}_{11}(t)))}{\hat{j}_{12}(t)}$ &  \\ 
 & $+g^2(t)[4 d_i^2-4 d_i \tau_{\mathds{D}}-4 \Delta_{\mathds{D}}+\tau_{\mathds{D}}^2] -4\Delta_{\mathds{D}} g'(t) $ & $+4 \Delta_{\mathds{D}} \hat{j}_{11}'(t)+\frac{d_u^2 \hat{j}_{12}'(t)}{\hat{j}_{12}(t)^2} $ &$\geq 0$ \\ 
 \hline
 \hline
 $k_m^2$ &  $\min_i  \left \{  \frac{(2 d_i-\tau_{\mathds{D}}) g(t) + \sigma_{\mathds{D\hat{J}}}}{2 \Delta_{\mathds{D}}} \right\} $  & $\min_{i,j\neq i}  \left \{  \frac{(2 d_i-\tau_{\mathds{D}}) g(t) + \sigma_{\mathds{D\hat{J}}}+d_i \frac{d}{dt} \log[\hat{j}_{ij}]}{2 \Delta_{\mathds{D}}} \right\} $ & --  \\
\hline
\end{tabular} 
 \caption{Turing conditions for a two-component system with isotropic growth. The middle column summarizes the conditions for a constant homogeneous state in Part 2 of this series. The right column presents the corrections for the time-dependent homogeneous state. $\tau$ and $\Delta$ refer to the trace and determinant of the matrix in the subscript, either $\sigma$ , the diagonal diffusion matrix $\mathds{D}$, or the Jacobian evaluated in the homogeneous state $\mathds {\hat{J}}$, depending on time.}\label{tab:conditions}
 \end{table*}

\section{A study case: linear growth/shrinking of the Brusselator}\label{sec:numeric}
Let us consider as an illustrative example the case where the growth/shrinkage is linear $l(t)=l(0)(1+ r t)$, where the growth rate is $g(t)=r/(1+ r t)$ . As we show in Part 1 of this work, the homogeneous state of linear growth ($r>0$) and shrinkage ($r<0$) changes in time and, in the first case, slowly tends to the point concentration fixed $ \mathbf{c}_0$, while in the second it slowly moves away from such value \cite{ledesma2023turingA}.

The Brusselator is given by

\begin{equation}\label{eq:bruselas1}
 \mathbf{f}(\mathbf{c})=(A-Bc_u-c_u +c_u^2 c_v, Bc_u -c_u^2 c_v)^T.
\end{equation}
The fixed point of the isolated reaction is in $\mathbf{c}_0=(A,B/A)^T$, and the Jacobian and diffusion matrix of the fixed domain problem in \eqref{eq:matrices} (see Ref. \cite{ledesma2020eckhaus}) are

\begin{equation}
\mathds{J}=\left(\begin{array}{cc}
-1+B & A^2 \\ 
-B & -A^2
\end{array}\right)   \mbox{ and }  
\mathds{D}=\left(\begin{array}{cc}
\sigma & 0\\ 
0 & 1
\end{array}\right).
\end{equation}

The linear approximation for the value of the homogeneous state if $|r|$ is relatively small is given by Eq. \eqref{eq:homoanalitic}, and the results are too long to put on one page. However, it is a useful approximation when explicit closed conditions are needed for the emergence of patterns. However, to avoid the approximation here and focus only on the hipotheses for the Turing conditions, for this work, we directly use the steady state obtained from the direct numerical solution of eq. \eqref{eq:homoanalitic}. Furthermore, from now on, to focus only on the effect of distance growth to the bifurcation, we will set $A=1$ and $\sigma=0.1$, which gives a critical value of the wavenumber and bifurcation parameters as $k_c = \sqrt{A/\sqrt{\sigma}}$ and $B_T=(1+A \sqrt{\sigma})^2$, respectively.

\begin{figure}[hbtp]
\centering
\includegraphics[width=0.75 \textwidth]{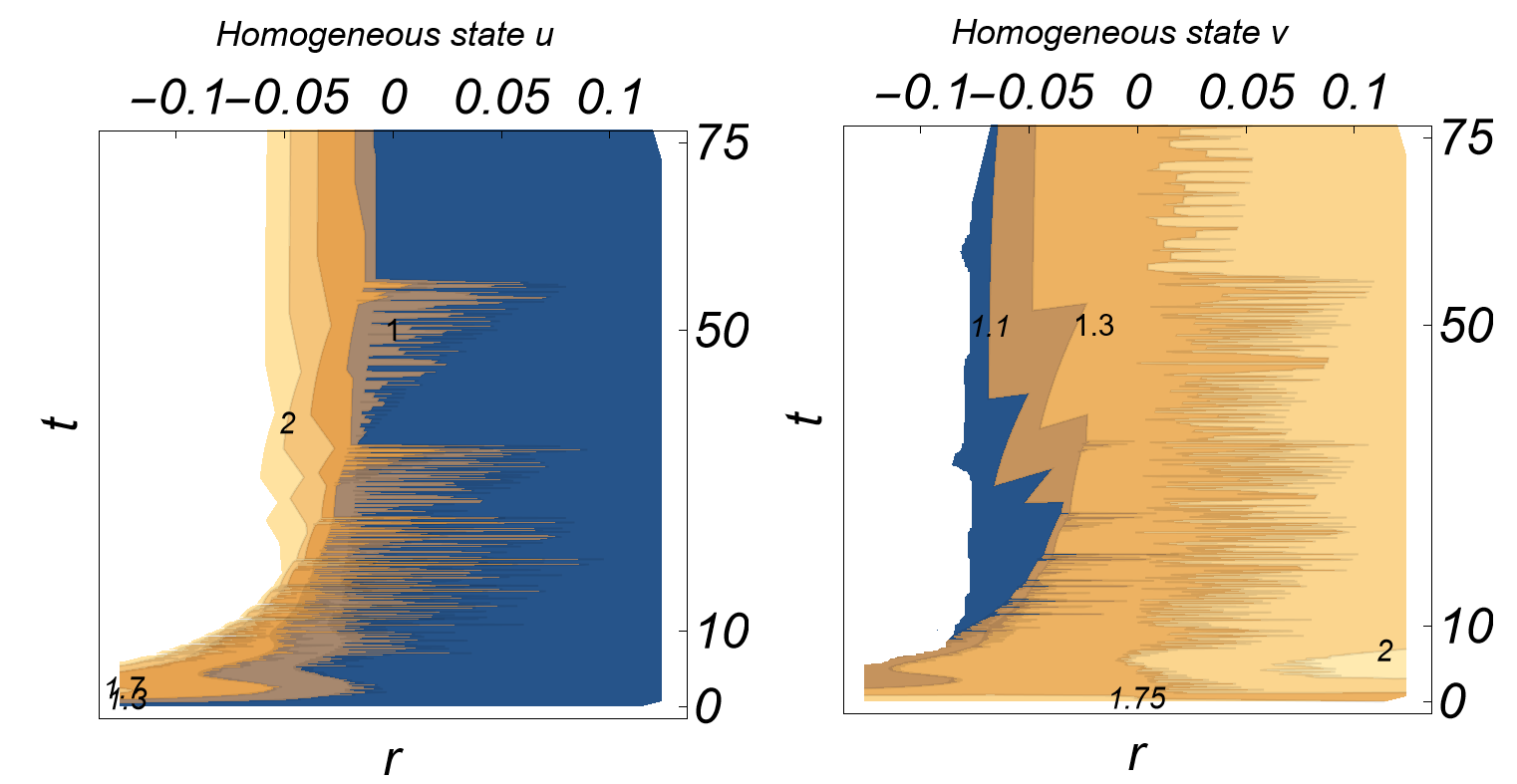}
\caption{Homogeneous state for both concentrations as a function of the parameter $r$ and time. We set the value $B=1.75$. The maximum time for simulation is chosen as the time needed to change the original size ten times. \label{fig:homogeneous}}
\end{figure}

In Fig \ref{fig:homogeneous}, we plot the value of the concentrations of the homogeneous state resulting from numerically solving the equation \eqref{eq:homoanalitic} for the value of $B=1.75$ between the times $t=0$ and $t_{max}$. We have used as initial condition $\mathbf{c}_s(0)=\mathbf{c}_0$. The final time for each value of $r$ is chosen as the time required for the domain to reach ten times its original size (growth $r>0$), or decrease ten times its original size (shrinkage, $r>0$ ). In this Fig \ref{fig:homogeneous},  we corroborate that in the case of growth, the value of homogeneous concentrations tends to its fixed point value, and in the case of shrinkage, it moves away from it as time progresses \cite{ledesma2023turingA} . Note also that there are rapid variations in the central area of this homogeneous concentration graph. These changes in turn can lead to rapid changes in the Turing region.

\begin{figure}[hbtp]
\centering
\includegraphics[width=0.75 \textwidth]{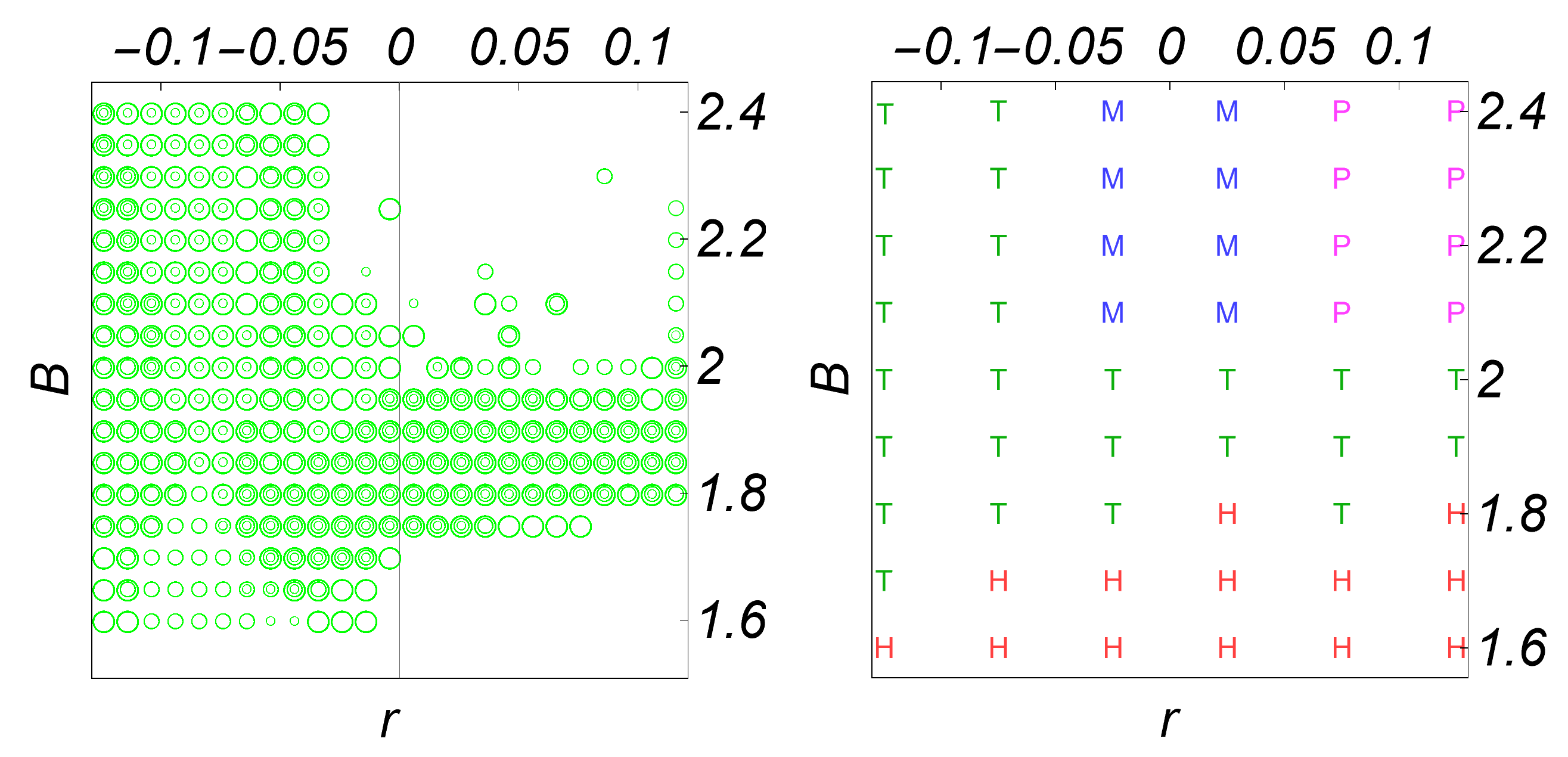}
\caption{Turing conditions for linear growth/shrinkage. Left.- Predictions of our scheme for three different moments represented by the size of the green circle. Right) Results of our numerical simulations with H,T,M and P representing homogeneous, Turing, mixed-mode and time-periodic numerical solutions, respectively. \label{fig:corroboration}}
\end{figure}

Now, knowing the homogeneous state, we can evaluate the Turing conditions deduced by us and summarized in the right column of Tab \ref{tab:conditions}. These conditions depend on time. To take this dependence into account, in Fig. \ref{fig:corroboration}.a, we have plotted three different circles, whose different diameters reflect the time in which the Turing conditions apply. Thus we evaluate the conditions at times $t_{max}/3$ (small circle), $t_{max}/2$ (medium circle) and $t_{max}$ (large circle), respectively. In this way, three concentric circles reflect that a pattern is predicted during almost the entire process, and two or one circle would mean a Turing pattern that appears/disappears over time.

In Fig. \ref{fig:corroboration}.b, we show the results of our numerical simulations performed in Comsol Multiphysics of the RDD system at \eqref{eq:system} for the Brusselator system at \eqref{eq:bruselas1} using linear growth. The simulations are performed in a fixed computational domain with 100 equidistant vertices with a simulation time $t_{max}$ calculated as the time it takes for the system to grow/shrink ten times the original size, depending on whether it is growing/shrinking, \emph{i.e.}, $t_{max}=9/r$ or $-9/10 r$, respectively.  The initial domain size is calculated using as reference the bifurcation wavenumber as $l (0)=2n\pi/k_c$, with $n$ equal to 3 or 19 for $r>0$ and $r<0$, respectively. We have used periodic boundary conditions and random disturbances of 10\% of the value of the initial concentration of $\mathbf{\hat{c}}_0$.

In Fig. \ref{fig:corroboration}.b we show with the letters H, T, M and P the numerical solutions corresponding to homogeneous, Turing, mixed-mode and only periodic solutions in time, respectively. As we explained in Part 2 of this series, homogeneous solutions are characterized by a low amplitude and a tendency to conserve wavenumber in the actual domain; Turing patterns have a medium amplitude and their spatial oscillations occur around a more or less fixed concentration; in the case of mixed mode spatial patterns it differs from Turing patterns in that they oscillate around a limit cycle and, finally, periodic solutions consist of limit cycles at each point without any predominant wave number in the domain.

As can be seen in figure \ref{fig:corroboration}, our theory allows us to predict that Turing patterns occur more broadly for domains that are shrinking and, in contrast, for growth they retain the same trend in the values of parameters than those where patterns occur in a fixed domain ($r=0$). Therefore, our scheme allows finding Turing patterns in growing domains even when the steady state changes over time.

\begin{figure*}[hbtp]
\centering
\includegraphics[width=0.9 \textwidth]{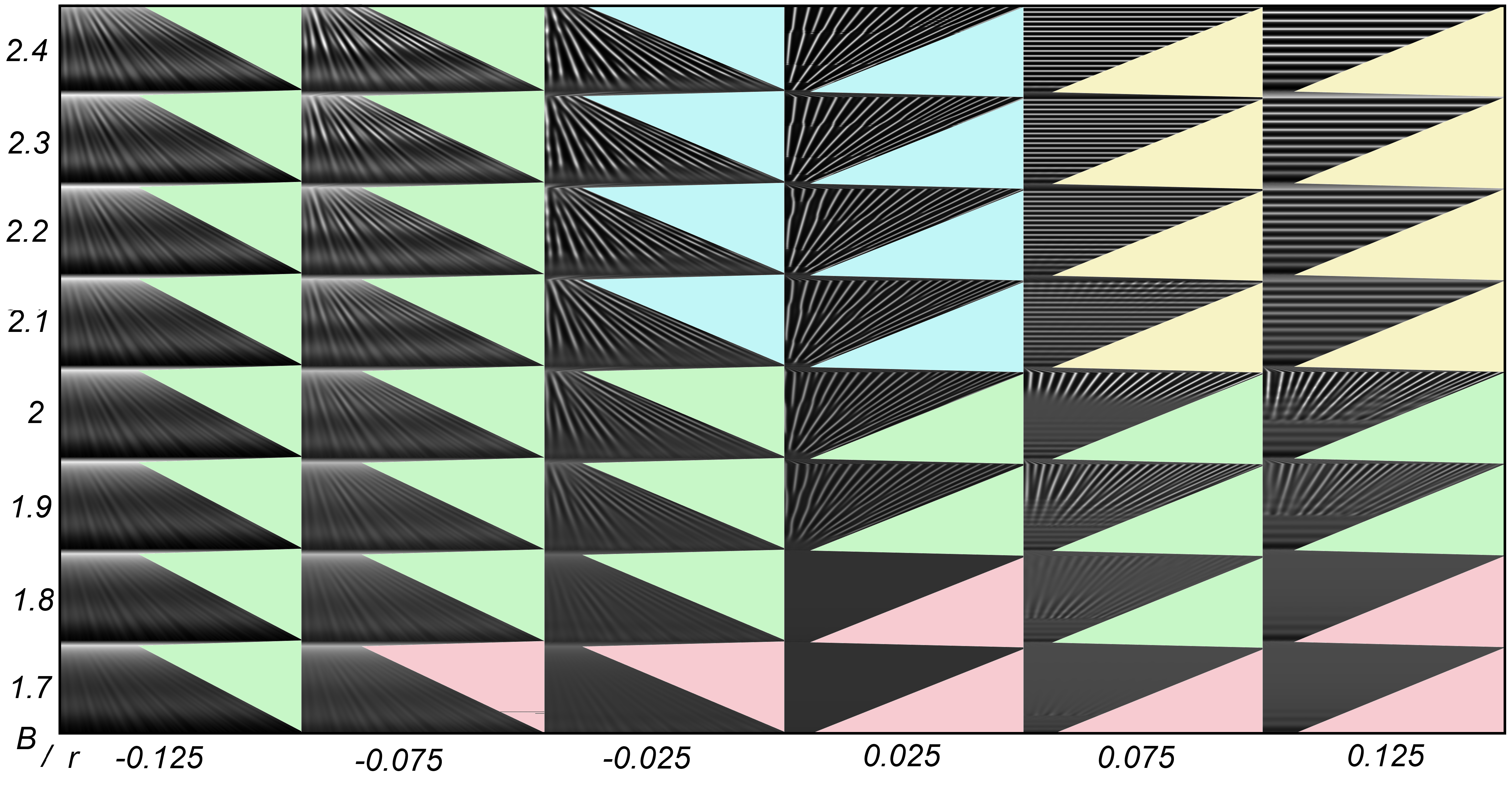}
\caption{Spatiotemporal maps of the numerical solutions presented in Fig. \ref{fig:corroboration}.a \label{fig:spacemap}}
\end{figure*}

Also in Fig. \ref{fig:spacemap} we observe that the zone over the Turing region presents both mixed-mode solutions and temporally periodic solutions without a spatial pattern. The characteristics of all these solutions can be better observed in the spatiotemporal maps presented in Fig. \ref{fig:spacemap}. These maps replicate the points in Fig. \ref{fig:corroboration}.b and show the qualitative changes between the different types of solutions. It should be noted that these spatial maps are not on a single spatial or color scale and are presented only to illustrate qualitative differences.

\begin{figure}[hbtp]
\centering
\includegraphics[width=0.75 \textwidth]{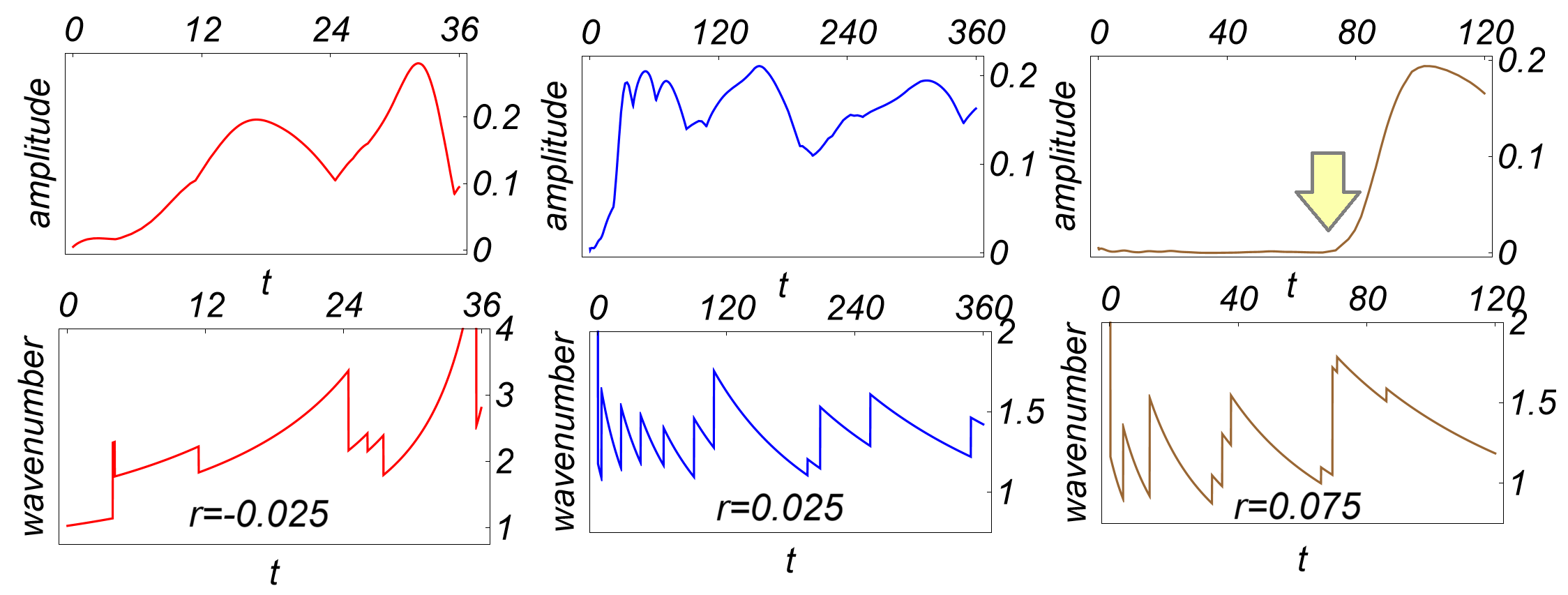}
\caption{Time-averaged wavenumber and amplitude for three different values of $r$ given in the inset. $B=2.0$ was set. Note the sudden appearance of pattern in the latter case. \label{fig:apparition}}
\end{figure}

The strictly temporal character of the conditions for the appearance of Turing patterns can be exemplified, for example, in the region close to $r \approx 0.075$ and $B\approx 2$ where, as illustrated in Fig. \ref{fig:spacemap}, A Turing pattern is not initially predicted, until later times. This can be corroborated numerically in Fig. \ref{fig:apparition}, where we have plotted the behavior of the wavenumber and the amplitude of the pattern for three different parameters. As observed in the last case, the amplitude of the Turing pattern is zero and only occurs until a later time. This suggests the role that time has in the Turing conditions as a possible bifurcation parameter and which will be studied later in this series of works.

\begin{figure}[hbtp]
\centering
\includegraphics[width=0.75 \textwidth]{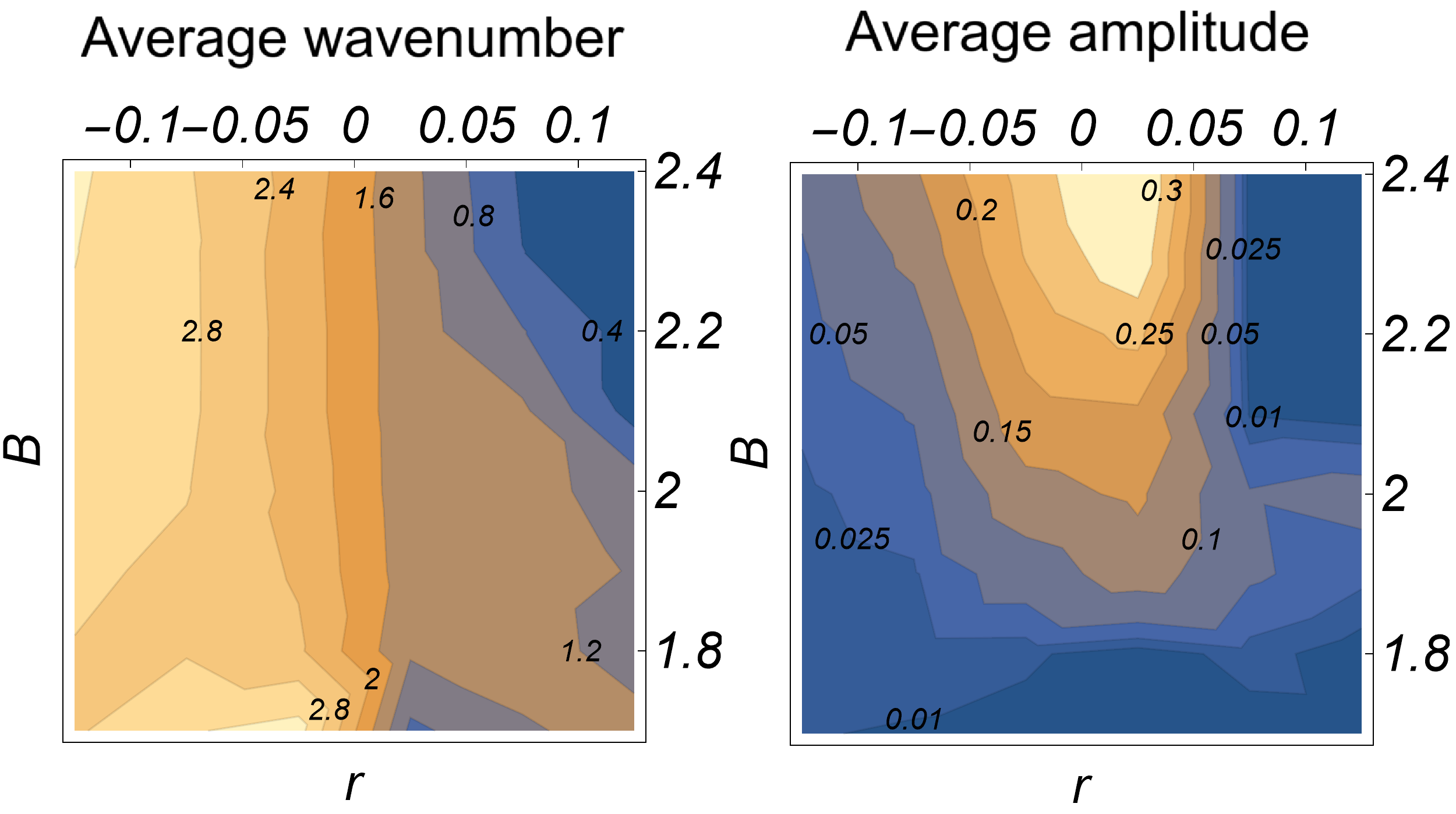}
\caption{Time averaged wavenumber and amplitude of all the numerical solutions on Fig. \ref{fig:corroboration}.a. \label{fig:features}}
\end{figure}

Finally, in Fig. \ref{fig:features}, we show some average characteristics of the numerical solutions found. In Fig. \ref{fig:features}.a, we find that the wavenumber in the actual domain, as in the exponential case, depends mainly on the growth parameter $r$. Therefore, for growing, the wavenumbers are smaller than for the shrinking case. This manifests the tendency of evolving-domain systems to have wave numbers at the bottom/upper part of the instability range for growth/shrinkage, respectively. Therefore, it seems that this feature is a characteristic of dilution itself rather than a specific type of growth. On the other hand, in Fig. In fig. \ref{fig:features}.b, we show the amplitude averaged over time. This figure corroborates that the region predicted for the Turing structures does indeed have a profile similar to that of Fig. \ref{fig:corroboration}.a predicted by us. It also shows that the periodic solutions in the upper right part of \ref{fig:corroboration}.b arise due to the impossibility of maintaining a non-zero amplitude for values of $r \gtrsim 0.05$.

\section{Discusion and conclusions}
In this work we have generalized the idea presented in Part 2 of this series of understanding the conditions for the formation of Turing patterns from the qualitative changes of a potential function of the linearized RDD problem. This has allowed us to hypothesize possible Turing conditions for domains that grow/decrease isotropically throughout their domain, and whose homogeneous state may depend on time.

These hypotheses were tested against the Brusselator-type RDD system with linear growth giving good comparisons between our predictions and the numerical simulations. These results, in addition to being evidence for our predictions, allows us to conclude that in linear growth the Turing region widens for shrinkage and changes slowly in the case of growth compared to a fixed domain system.

We also corroborate that the Turing conditions, for the case in which the homogeneous state depends on time also can depend on time, and therefore it may be the case that a pattern appears or disappears as the domain evolves.

As in the case of exponential,  we corroborate that for linear  growth/shrinkage, the wave number on average is smaller/larger than for the case of a fixed domain, respectively, demonstrating that this behavior of the wave number is more of an inherent property of a diluted system that changes in size, rather than the particular type of growth. We also find that the average amplitude of the patterns in the case of linear growth is similar to those obtained with exponential growth studied before.

We conclude that it is still necessary to compare the results of this work with the predictions made in previous works on Turing conditions for increasing domains, for example \cite{van2021turing,madzvamuse2010stability}. However, we believe that the theory presented in this series offers a panoramic vision that allows us to understand the formation of spatial patterns in a general way.

\bibliography{biblio1}

\end{document}